# A Joint Processing Strategy for Image Quality Improvement in 3D Digital Subtraction Angiography

Xiaoxuan Zhang[1], Xiao Jiang[2], Matthew Tivnan[2], J. Webster Stayman[2], and Grace J. Gang[1]

[1] Hospital of the University of Pennsylvania Department of Radiology
[2] Johns Hopkins University Department of Biomedical Engineering

**Abstract** Three-dimensional digital subtraction angiography (3D-DSA) is a widely adopted technique for clinical evaluation of contrast-enhanced vasculatures. The distribution of a contrast agent such as iodine is often estimated via temporal subtraction. Advancements in spectral imaging technologies such as photon counting detectors offer new opportunities to improve DSA image quality. In this work, we propose a novel joint processing strategy to achieve an iodine image using two-bin spectral measurements from a photon counting detector acquired both the pre- and post-contrast injection. Simulation studies were performed using a digital phantom with iodine-enhanced vessels. The proposed method was compared with temporal subtraction and conventional spectral imaging using just the post-contrast measurements. Imaging performance was evaluated in terms of noise-resolution tradeoffs. Preliminary findings have shown measurably improved image quality given by joint processing, reducing noise by 40% and 70% compared to temporal subtraction and conventional spectral imaging using an energy-integrating detector, respectively.

## 1 Introduction

Three-dimensional digital subtraction angiography (3D-DSA) has been widely used for the diagnosis of vascular diseases in a range of clinical applications.[1,2] Conventionally, 3D-DSA relies on temporal subtraction, where cone-beam CT (CBCT) acquisitions pre- and post-contrast injection are subtracted to isolate iodine distribution in the vasculature.

With the increasing availability of spectral hardware, efforts to introduce spectral imaging for interventional applications are quickly emerging.[3] Spectral imaging has been previously proposed for 2D DSA as an alternative to temporal subtraction to mitigate patient motion. A post-contrast only acquisition using spectral measurements allow iodine distribution to be computed using material decomposition algorithms. However, significant noise amplification from material decomposition impeded clinical adoption.

Recent work[4] has proposed a joint processing strategy using spectral measurements acquired both pre- and post-contrast injection to improve the noise performance 2D DSA image estimates. Such strategy leverages both the temporal and spectral information and can yield DSA images that outperform conventional temporal-only and spectral-only processing. This work seeks to apply similar strategies to 3D-DSA imaging. We focus on spectral measurements using a photon counting detector (PCD) due to its superior spectral separability and spatial resolution compared to other spectral technologies. We envision that the low image noise and high spatial resolution enabled by this strategy can bring the most benefit to the visualization of small, low contrast targets like small vessels.

## 2 Methods

2.1. Theoretical methods

We propose a novel joint processing strategy for 3D-DSA which leverages both temporal and energy information for image quality improvement. Projection data are acquired using a spectral imaging system before and after contrast injection. Measurements in two spectral channels are obtained for each acquisition, yielding a total of four measurements. Under the assumption of minimal patient motion (valid for intubated patients in anatomical sites without involuntary motion like cardiac and breathing), we can formulate a general forward model that relates a three-material object (water, bone/calcium, iodine) to the four measurements as follows:

$$\bar{y}_1(l) = \mathbf{S}_1 \exp(-\mathbf{Q}^{H_2O}\mathbf{A}\rho^{H_2O} - \mathbf{Q}^{Ca}\mathbf{A}\rho^{Ca}) \quad (1)$$

$$\bar{y}_2(l) = \mathbf{S}_2 \exp(-\mathbf{Q}^{H_2O}\mathbf{A}\rho^{H_2O} - \mathbf{Q}^{Ca}\mathbf{A}\rho^{Ca}) \quad (2)$$

$$\bar{y}_3(l) = \mathbf{S}_1 \exp(-\mathbf{Q}^{H_2O}\mathbf{A}\rho^{H_2O} - \mathbf{Q}^{Ca}\mathbf{A}\rho^{Ca} - \mathbf{Q}^{I}\mathbf{A}\rho^{I}) \quad (3)$$

$$\bar{y}_4(l) = \mathbf{S}_2 \exp(-\mathbf{Q}^{H_2O}\mathbf{A}\rho^{H_2O} - \mathbf{Q}^{Ca}\mathbf{A}\rho^{Ca} - \mathbf{Q}^{I}\mathbf{A}\rho^{I}) \quad (4)$$

where $\{\bar{y}_1, \bar{y}_2\}$ denotes the pre-contrast mean measurements containing water and bone/calcium, $\{\bar{y}_3, \bar{y}_4\}$ denotes the post-contrast mean measurements containing water, calcium, and iodine, $\mathbf{S}_1$ and $\mathbf{S}_2$ models the spectral sensitivities of each spectral channel, $\mathbf{Q}$ contains the mass attenuation coefficients for each basis material, $\mathbf{A}$ denotes the system matrix, and $\rho$ represents material densities. We can alternatively combine the four equations into a single equation to facilitate joint estimation:

$$\begin{bmatrix}\bar{y}_1\\\bar{y}_2\\\bar{y}_3\\\bar{y}_4\end{bmatrix} = \begin{bmatrix}\mathbf{S}_1\\&\mathbf{S}_2\\\mathbf{S}_1\\&&\mathbf{S}_2\end{bmatrix} \exp\left(-\begin{bmatrix}\mathbf{Q}\\\mathbf{Q}\end{bmatrix}\begin{bmatrix}\mathbf{I}\\&\mathbf{I}&\mathbf{0}\\\mathbf{I}\\&&\mathbf{I}\end{bmatrix}\begin{bmatrix}\mathbf{A}\\&\mathbf{A}\\&&\mathbf{A}\end{bmatrix}\begin{bmatrix}\rho^{H_2O}\\\rho^{Ca}\\\rho^{I}\end{bmatrix}\right) \quad (5)$$

Here, a masking matrix composed of identity matrices and zeros selectively zero out the pre-contrast measurements associated with iodine.

*2.1.1. Model-based one-step decomposition/reconstruction*

This form of forward model (linear-exponential-linear) can be solved by an optimization algorithm using a separable paraboloidal surrogate approach previously developed by the authors.[5] Assuming the measurements follow a multivariate Gaussian distribution, $y \sim \mathcal{N}(\bar{y}, \Sigma)$, the material density, $\rho$, can be solved using a nonlinear weighted least squares objective:

$$\hat{\rho} = \arg\min_{\rho} \Phi(\rho, y), \text{ where} \quad (6)$$

$$\Phi(\rho, y) = (y - \bar{y}(\rho))^T \Sigma_y^{-1}(y - \bar{y}(\rho)) + \beta R(\rho) \quad (7)$$

A quadratic penalty, $R$, is adopted in this work. We refer to this algorithm as the model-based one-step decpomposition (MB-OSD). The densities of water, calcium, and iodine were initialized with zeros and iteratively solved using 600 iterations of the separable quadratic surrogate algorithm with 10 ordered subsets[6] and Nesterov's acceleration.[7]

*2.1.2. Model-based projection-domain decomposition*

Alternatively, a more memory efficient and faster algorithm involves performing a projection-domain decomposition followed by an analytic or model-based reconstruction (MB-PDD). Instead of physical density $\rho$, we now estimate the material density line integral, $l = \mathbf{A}\rho$. The forward model in Eq. (5) can be modified to:

$$\begin{bmatrix} \bar{y}_1 \\ \bar{y}_2 \\ \bar{y}_3 \\ \bar{y}_4 \end{bmatrix} = \begin{bmatrix} S_1 \\ & S_2 \\ S_1 \\ & S_2 \end{bmatrix} \exp\left(-\begin{bmatrix} \mathbf{Q} \\ \mathbf{Q} \end{bmatrix}\begin{bmatrix} \mathbf{I} \\ & \mathbf{I} & \mathbf{0} \\ \mathbf{I} \\ & & \mathbf{I} \\ & & \mathbf{I} \end{bmatrix}\begin{bmatrix} l^{H_2O} \\ l^{Ca} \\ l^I \end{bmatrix}\right) \quad (8)$$

We can write down a similar objective function as above based on $l$:

$$\Phi(l, y) = (y - \bar{y}(l))^T \Sigma_y^{-1}(y - \bar{y}(l)) \quad (9)$$

$$\hat{l} = \arg\min_{l} \Phi(l, y) \quad (10)$$

Note that no additional regularization is included in this objective. We instead rely on smoothing in the volumetric reconstruction (e.g., apodization filter in FDK or regularization in model-based iterative reconstruction).

We adopted a modified Newton's method to solve this objective, where a proximal regularization term was added to the diagonal of the Hessian to improve conditioning of the inversion process. We adoped zero initialization and applied 50 iterations of optimizer. Volumetric reconstruction of the estimated material density line integrals may be performed using either analytical or iterative algoirthms. In this work, we adopted filtered back-projeciton (FBP) with a Hann apodization filter with different cutoff frequencies.

*2.2. Experimental methods*
*2.2.1. Phantom*

A digital phantom (Figure 1) was constructed for data generation, which includes a 1000 mg/mL water background, a 4.0 mm-thick 200 mg/mL Calcium ring, and 10 mg/mL iodine-enhanced vessels of diameter ranging from 0.3 mm to 4.0 mm. The phantom is intended for evaluation of high-resolution imaging of small targets such as intracranial vessels. Pre-contrast imaging involved only the water background and the calcium ring, while post-contrast imaging included the enhanced vessels.

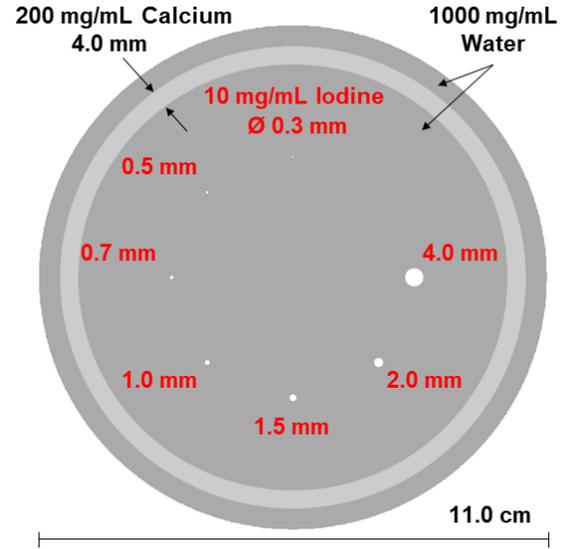

Figure 1. Digital phantom with a water background, a calcium ring, and iodine-enhanced, cylindrical vessels of varying diameters.

*2.2.2. Comparison studies*

We simulated four 3D-DSA techniques for comparison:

(1) Pre- and post-contrast measurements generated using an energy integrating detector (EID) with a 600-μm CsI scintillator. These measurements emulate data from the current standard DSA using temporal subtraction and what might be used in a 3D-DSA interventional imaging system (e.g., C-arm);

(2) Pre- and post-contrast measurements generated using a single energy bin (10–150 keV) from a photon counting detector with 750-μm CdTe. This dataset was used to evaluate performance of temporal subtraction with a PCD;

(3) Post-contrast measurements only from a PCD. Two energy thresholds were applied (40 keV and 60 keV) to obtain data from three energy bins. The PCD was modeled with "perfect" energy resolution. This data provided an example of "conventional" spectral imaging in 3D-DSA, where only energy information was leveraged for material decomposition. This technique is evaluated at both 1x and 2x the total dose for a fair comparison to techniques using both the pre- and post- measurements; and

(4) Pre- and post-contrast measurements from the PCD with two energy bins. This dataset contained both temporal

and energy information and was used to evaluate the proposed joint processing strategy. The energy threshold is set to 50 keV.

*2.2.3. Simulation*

Projection data for 3D imaging (1080 × 5 pixels, 0.154 mm pixel pitch, 360 views uniformly distributed over 360°) were generated for a system geometry with a source-to-axis distance of 800 mm and a source-to-detector distance of 1100 mm. The x-ray spectrum was generated calculated using Spektr[8] at 100 kVp with 0.25 mm Aluminum as the intrinsic filtration. The nominal tube current-time product was set to 1.0 mAs per projection. Poisson noise was added to all measurements, and detector blur measured from a physical energy integrating detector was applied to data from the EID. Blur was assumed to be negligible for the PCD. Volumetric images were reconstructed at 0.1 mm isotropic voxel size.

In temporal subtraction, the iodine distribution was obtained by subtracting two single energy reconstructions. The scale and units are therefore mismatched from physical density obtained from spectral decomposition. To enable direct quantitative comparisons, an "oracle" scaling factor computed from noiseless measurements was applied to convert attenuation coefficient differences to iodine density. In conventional spectral imaging, the density line integrals were estimated via model-based projection domain decomposition considering three energy sensitive channels and only post-contrast measurements.

For quantitative comparison of methods, the noise-resolution trade-off in the iodine basis image was evaluated by varying the cutoff frequency of the Hann filter. A similar curve can be generated by tuning the regularization strength for the one-step decomposition and will be the subject of future work. Resolution was characterized by the full width at half maximum (FWHM) of the 0.3 mm diameter vessel measured in a noiseless reconstruction. Noise was measured within a 21 × 21 × 1 region of interest at the center of the 4.0 mm diameter vessel over 10 repeated reconstructions with different noise realizations.

## 3 Results

Figure 2 shows the noise-resolution curves for conventional spectral imaging, temporal subtraction, and joint processing with model-based projection domain decomposition. Across all resolutions, conventional spectral imaging exhibits higher noise levels, even when the exposure is increased by a factor of two. Temporal subtraction using the EID provides reduced noise but also exhibits a limited improvement in spatial resolution compared to other methods – i.e., the resolution improvement stopped around 0.45 mm and increasing cutoff frequency did not improve resolution. In comparison, temporal subtraction using the PCD offers further opportunity to improve resolution, as well as ~25% reduction in noise at matched resolution. Joint processing with MB-PDD exhibited the best noise-resolution trade-off overall, permitting a ~12% noise reduction compared to temporal subtraction using the same PCD.

Figures 3(a-e) shows example decomposed iodine images of the 0.3 mm diameter vessel at a matched resolution as indicated by the black dotted vertical line in Figure 2. A high level of noise is present in Figures 3(a-b), obscuring the small vessel target. By contrast, the vessel can be easily differentiated from background noise in Figures 3(c-e), while the magnitude of noise is noticeably lower in (d) and (e).

Figure 3(f) demonstrates a preliminary model-based one-step decomposition result, which exhibits higher resolution (~0.3 mm) and markedly reduced noise compared to other methods. We conjecture that is due to the nonnegativity constraint clipping the noise at 0. While the strong noise suppression may be desirable for angiography applications, the images tend to appear "patchy" and low contrast features may be suppressed. Further investigation is underway where non-negativity constraint is turned off.

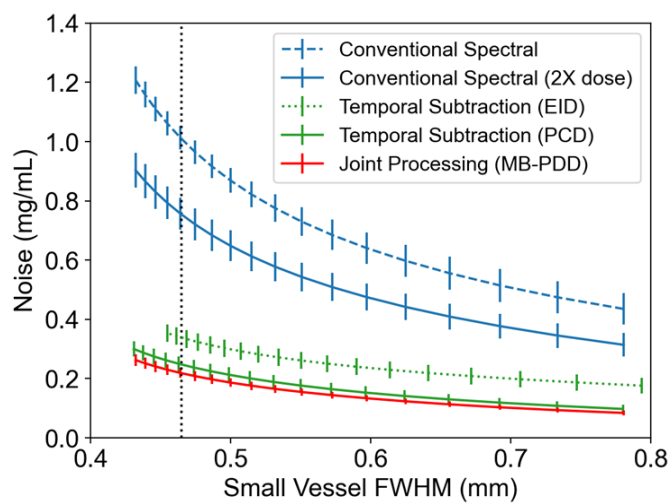

Figure 2. Noise-resolution characterization in iodine basis images estimated through conventional spectral imaging, temporal subtraction, and joint processing with varying cutoff frequencies for FBP reconstruction. Example images at matched resolution (dotted vertical line) are shown in Figure 3.

## 4 Discussion and Conclusion

This work demonstrated the capability of a joint processing strategy for improved noise and resolution performance in 3D-DSA. Leveraging both temporal and energy information, model-based decomposition approaches was shown to yield better noise-resolution tradeoffs compared to temporal subtraction or material decomposition based solely on multi-energy data.

Ongoing work includes spectral system optimization of the joint processing strategy. Material decomposition algorithms will be further optimized, including a comprehensive regularization parameter sweep of the one-step method and investigations of the behaviors of the nonnegativity constraint. Modeling of non-ideal effects in PCD including charge sharing and pulse pileup, and integration of registration algorithms to account for tissue deformation during image acquisition are also subjects of future work.

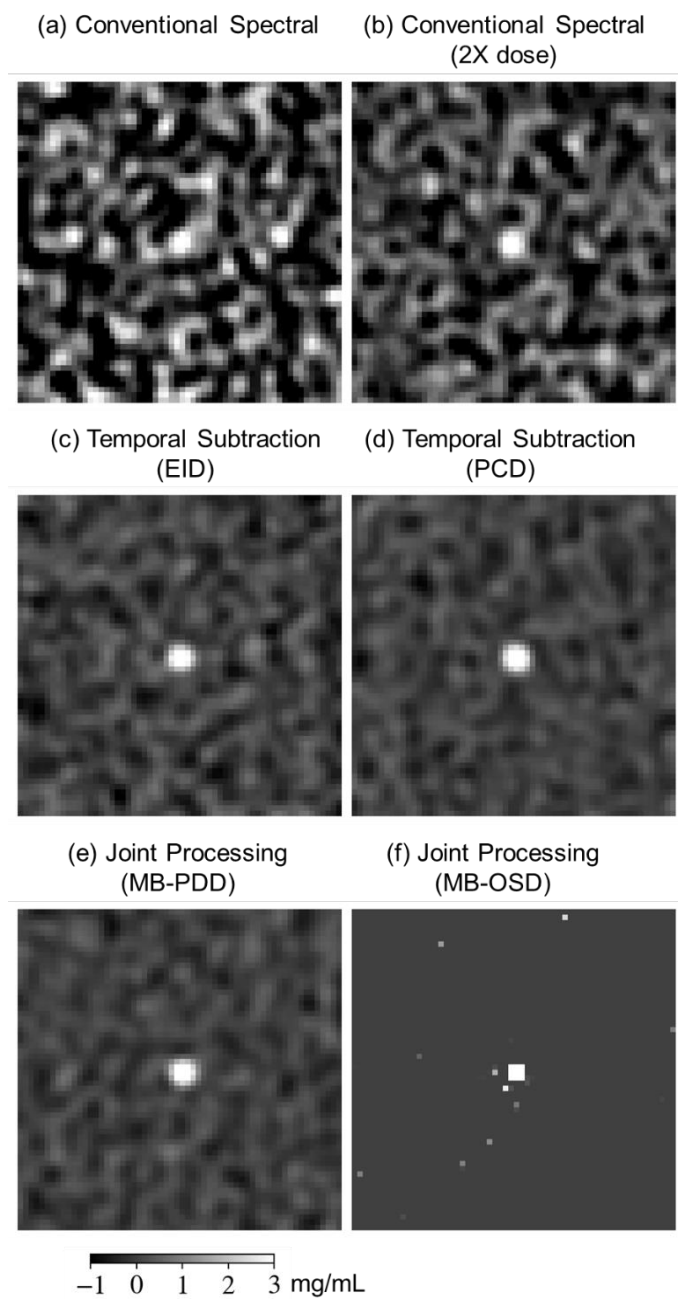

Figure 3. Example axial slice of the iodine basis images of the 0.3 mm diameter vessel estimated through (a-b) conventional spectral imaging at two dose levels, (c-d) temporal subtraction using the EID and the PCD, and (e-f) joint processing with MB-PDD and MB-OSD. The resolution was matched at 0.465 mm for (a-e) (dotted vertical line in Figure 2), and the examples were given by the mean noise level of 10 repeats. The same display window setting was applied to all images.